\documentclass{emulateapj}

\newcommand{\msun}{$M_\odot$}
\newcommand{\etot}{$E_{tot}$}

\newcommand{\fe}{\mathrm{[Fe/H]}}

\newcommand{\kms} {$\mathrm{km \ s^{-1}}$}
\newcommand{\lz}{$L_z$}
\newcommand{\lp}{$L_{\perp}$}
\newcommand{\rgc}{$R_{gc}$}
\newcommand{\meanlz}{$\langle L_z \rangle$}

\bibpunct{(}{)}{,}{a}{}{;}

\shorttitle{Fashionably Late?}
\shortauthors{Morrison et al.}

\begin{document}

  \title{Fashionably Late? Building up the Milky Way's Inner Halo}

  \author{Heather L. Morrison}
  \affil{Department of Astronomy, Case Western Reserve University, Cleveland, OH 44106 \email{heather@vegemite.case.edu}}
  \author{Amina Helmi}
  \affil{Kapteyn Astronomical Institute, University of Groningen, PO Box 800, 9700 AV Groningen, the Netherlands \email{ahelmi@astro.rug.nl}}
  \author{Jiayang Sun, Peng Liu and Rongfang Gu}
  \affil{Department of Statistics, Case Western Reserve University,
  Cleveland, OH, 44106
  \email{jiayang.sun@case.edu,peng.liu@case.edu,rongfang.gu@case.edu}}
  \author{John E. Norris}
\affil{Research School of Astronomy and Astrophysics, The Australian
  National University, Mt Stromlo Observatory, Cotter Road,
Weston ACT 2611, Australia \email{jen@mso.anu.edu.au}}
  \author{Paul Harding}
  \affil{Department of Astronomy, Case Western Reserve University,
  Cleveland, OH 44106 \email{harding@dropbear.case.edu}}
\author{T.D. Kinman}
\affil{National Optical Astronomy Observatories, PO Box 26732, Tucson
  AZ 85726 \email{tkinman@noao.edu}} 
  \author{Amanda A. Kepley\footnote{Current address: Department of Astronomy, University of Virginia, Charlottesville, VA 22904-4325} }
  \affil{ Department of Astronomy, University of Wisconsin--Madison, 475 North Charter Street, Madison, WI 53706 \email{kepley@astro.wisc.edu}}
  \author{Kenneth C. Freeman}
\affil{Research School of Astronomy and Astrophysics, The Australian
  National University, Mt Stromlo Observatory, Cotter Road,
Weston ACT 2611, Australia \email{kcf@mso.anu.edu.au}}
  \author{Mary Williams}
\affil{Research School of Astronomy and Astrophysics, The Australian
  National University, Mt Stromlo Observatory, Cotter Road,
Weston ACT 2611, Australia  and
Astrophysikalisches Institut Postdam, An der Sternwarte 16, 
D-14482 Potsdam, Germany \email{mary@aip.de} }
 
  \author{Jeffrey Van Duyne}
  \affil{Department of Astronomy, Yale University, P.O. Box 208121,
    New Haven, CT 06520 \email{vanduyne@astro.yale.edu}} 

\begin{abstract}

Using a sample of 246 metal-poor stars (RR Lyraes, red giants and RHB
stars) which is remarkable for the accuracy of its 6-D kinematical
data, we find, by examining the distribution of stellar orbital angular momenta, a new component for the local halo which has an axial
ratio $c/a \sim 0.2$, a similar flattening to the thick disk.  It has
a small prograde rotation but is supported by velocity anisotropy, and
contains more intermediate-metallicity stars (with $-1.5 < \fe <
-1.0$) than the rest of our sample. We suggest that this component was
formed quite late, during or after the formation of the disk. It
formed either from the gas that was accreted by the last major mergers
experienced by the Galaxy, or by dynamical friction of massive
infalling satellite(s) with the halo and possibly the stellar disk or
thick disk.

The remainder of the halo stars in our sample, which are less closely
confined to the disk plane, exhibit a clumpy distribution
in energy and angular momentum, suggesting that the early, chaotic
conditions under which the inner halo formed were not violent enough
to erase the record of their origins. The clumpy structure suggests
that a relatively small number of progenitors were responsible for
building up the inner halo, in line with theoretical expectations.

We find a difference in mean binding energy between the RR Lyrae
variables and the red giants in our sample, suggesting that more of
the RR Lyraes in the sample belong to the outer halo, and that the
outer halo may be somewhat younger, as first suggested by
\citet{sz}. We also find that the RR Lyrae mean rotation is more
negative than the red giants, which is consistent with the recent result of
\citet{daniela} that the outer halo has a retrograde rotation and with
the difference in kinematics seen between RR Lyraes and BHB stars by
\citet{kinman07}.

\end{abstract}


\section{Introduction: Reconstruction of halo history}

One of the central aims of the study of old populations in the Galaxy
is the reconstruction of its history from their properties today. We
will begin here by considering the different approaches to modeling
the structure of the halo. \citet{sz} took an important first step
toward our current understanding of the hierarchial build-up of
galaxies when they noted that ``the loosely bound clusters of the
outer halo have a broader range of age than the more tightly bound
clusters.'' Perhaps because of our taxonomic heritage, subsequent
stellar populations workers have tended to describe the complex
structure of the stellar halo in terms of two components.  For
example, in one of the first attempts at ``Armchair Cartography'' (the
use of a local sample of halo stars to reconstruct the halo density
distribution) \citet{jesper} inferred a two-component halo, with one
component quite flattened.  However, the study was limited by the
small sample and the relatively inaccurate distances and velocities
then available.

More global studies of the halo using kinematically unbiased samples
also found variants on this two-component halo. \citet{kinman65} and
\citet{hartwick} used RR Lyrae variables to show that the flattened
component dominates in the inner halo, and the spherical component in
the outer halo.  \citet{preston91} and \citet{kinsuntkraft} found a
similar spatial distribution using blue horizontal branch stars and RR
Lyraes. \citet{chibabeers} used a larger, local sample and confirmed
the results of previous studies. Most recently, \citet{daniela} used
more than 20,000 local stars from SDSS to further quantify this
``halo dichotomy''. They find that the outer halo is significantly
more metal-poor than the inner halo, and has a retrograde mean
rotation.

The influential work of \citet{zinn93} \citep[building on the earlier
work of][]{sz} developed the two-component halo idea further, using
globular cluster kinematics. Zinn subdivided the halo globular
clusters into different groups using horizontal branch morphology (as
a proxy for age) and found different spatial distributions and
kinematics for the two groups.  The Old Halo clusters in the inner
halo show a flattened distribution and have prograde rotation
(although much less than the disk's).  The Younger Halo is spherical
and has no net rotation.

With the exception of the work of \citet{jesper} and
\citet{chibabeers}, all of the works which derived a flattened inner
halo were based on samples of stars more distant than $\sim$1
kpc. However, \citet{preston91} noted a curious anomaly: when
they used their density model (including an inner halo with $c/a$=0.7
at the Sun) to predict the density of RR Lyraes in the solar neighborhood,
the model predictions fell seriously short: by a factor of at least
two, even though the local sample is unlikely to be complete at very
low latitudes.  Preston
et al.\ followed this fact to its logical conclusion and suggested
that the halo has an additional highly flattened component found only
very close to the plane. This paper will confirm the detection of the
highly flattened halo component with a new sample of halo stars which
have a median distance of 1 kpc.

How can we explain these results in the context of current models of
galaxy formation? There have been significant advances in recent years
which have developed a more nuanced picture of galaxy formation in a
number of areas. The concept which has changed least over the years is
that of disk formation: a smooth, dissipational collapse which
conserves the angular momentum of the infalling material seems the
only way to produce large galaxy disks like our own. Halo formation
was likely much messier. Our concepts here have changed more with
time, as the hierarchical paradigm for structure formation has
permeated the field: we now understand that accretion is fundamental
to the formation of the dark halo. Classical expectations are that
this accretion is violent in its early stages: the rapid variations in
the gravitational potential caused by early mergers should smooth out
any record of the discrete origins of early halo stars. We will see
below that this is not entirely correct.

Before the disk forms, halo stars
could be produced by the mergers of gas-rich objects \citep[the
``dissipative mergers'' of ][]{bekkichiba}. (Only minor accretions can
occur after the present-day disk forms, because otherwise the fragile
kinematically cold disk would be destroyed). The stellar halo can also
be built up by accretions of larger objects, helped by dynamical
friction (which was likely the way the progenitor of the globular
cluster $\omega$ Centauri arrived in the inner halo), or by the simpler
accretion of smaller satellites which did not feel dynamical
friction. We will see that there are opportunities for all of these
processes to build up a flattened inner halo.

The use of integrals such as energy and angular momentum, which are
conserved under some conditions as the Galaxy evolves, can be a
powerful tool for tracing the Galaxy's history. For example,
\citet{lb2} and \citet{kathrynpoles} showed how energy can be used to
identify stars with common origins, and
\citet{amina_nature,amina00,chibabeers} and \citet{amina06} considered
different components of angular momentum and energy.
\citet{amina_nature} illustrate the different information
available from study of stellar velocities and of angular momenta:
their Figure 2 shows that the debris from a disrupting satellite is
all found in the same place in the angular momentum diagram, but
that different streams from this satellite may be seen with distinct
velocities (moving towards and away from the plane, for example).

In this paper we analyse a sample of metal-poor stars from the solar
neighborhood which is remarkable for the accuracy of its kinematic
data. These high-quality data have well-determined orbital angular
momenta and energies, making them well-suited to the approach using
integrals. We find patterns in the distribution of energy and
angular momentum, never observed before, which we will relate to
different formation and evolution processes.  The location of our
sample (its stars have a median distance of 1 kpc) influences the
processes we can study. The solar neighborhood lies within the inner
halo\footnote{Contrast the Sun's \rgc\ of 8 kpc with distant halo stars at
more than 100 kpc \citep{clewley05}} but it is also centered on the
disk.  We will focus in this paper on processes which either form
stars in the inner halo or transfer stars into this region.  We will
seek to distinguish between stars formed very early in the inner
galaxy (its original inhabitants) and the fashionably late arrivals
which were accreted later from the outer halo. These could either have
orbits sufficiently eccentric to bring them into the inner halo or had
their orbits modified by dynamical friction on the parent satellite to
move them from the outer to the inner halo.

The plan of this paper is as follows.  In Section 2 we describe our
sample and investigate the properties of these stars in angular
momentum, energy and metallicity; and discuss the trends that we
discern.  In Section 3 we will relate our results to the work of
others, and in Section 4 we will discuss the different formation paths for
the inner halo, and how our data constrain them.

\section{Analysis of local halo sample}

\subsection{Sample}

Our sample is composed of red giants, red horizontal branch (RHB)
stars and RR Lyrae variables with [Fe/H]$<-1.0$ and distances less
than 2.5 kpc.  The red giants and RHB stars are all taken from the
sample of well-studied stars in \citet{att94}, and comprise all the
stars in that study with available velocity data.  Its RR Lyrae
variables include a large number with new, accurate radial velocity
measurements (van Duyne et al, in preparation) and a few with
velocities from the literature with similar or higher accuracy.  The
sample contains 246 stars, and can be accessed at
http://astronomy.case.edu/heather/fashlate.sample.

The sample has a large overlap with the ``combined data sample'' of
\citet{kepley07}; but we have restored 12 stars which they removed
from their sample because they were likely metal-weak thick disk
stars, in order to make the sample selection criteria clearer.  We
removed 22 giant stars which were in the Kepley et al sample but not
in the sample of \citet{att94}, as they have distances of
significantly lower quality. We have also added a few giants from
\citet{att94} which were not in the Kepley et al sample because they
had radial velocities measured after 2000, and so were not in the
catalog of \citet{beers00} on which it was based.

Our sample contains stars with accurate distances and well-quantified
distance errors, allowing us to calculate errors on derived quantities
like energy and angular momentum via a Monte-Carlo procedure.
There is a full description of our distance error calculation (which
used a Monte-Carlo procedure to take account of all the different
contributing errors) in Section 2.2.3 of \citet{kepley07}.   We
emphasise the importance of our small distance errors here: energy and
angular momentum are strongly dependent on distance errors since two
out of three components of velocity are obtained by multiplying the
proper motion by the distance.

Except for the three stars with new velocities, our sample is a
subset of the compilation of halo stars of \citet{beers00}. We have
improved the data in several ways. First, we
removed the effect of the large distance errors in the sample found by
\citet{kepley07} by limiting the giants and RHB stars in our sample to
those with very well-defined luminosities and metallicities, by using
only those from the sample of \citet{att94}.
The \citet{beers00} sample contains 736 stars with $\fe < -1.0$ and
distances less than 2.5 kpc which have proper motion measurements. 435
of these stars are red giants, RHB stars or RR Lyrae variables. We
estimate that 60\% of these stars have the well-determined distances
and well-quantified distance errors that are required for our study;
the other 40\% have data of significantly lower quality, either
because of the lack of a luminosity measurement which separates stars
below the horizontal branch from those on the horizontal branch, or
because the metallicity measurements are less accurate and not on a
consistent scale.  However, RR
Lyrae variables have large pulsational velocity amplitudes and so
require significant effort to constrain their systemic velocities; many
of the RR Lyraes in \citet{beers00} have large errors on their velocities. The
restriction of our sample to those with both good distance and
velocity measurements cuts the sample down to 246 stars\footnote{We
  have discarded one star, HD220127, which has large errors on its
  derived velocities using the distance from \citet{att94} and a
  significantly different distance in \citet{schuster06}.}. It has a
median distance of 1 kpc and a median distance error of 7\%  \citep[error
estimates for red giant and RHB distances are given in Table 1 of][]{kepley07}. 

Since our giant and RHB sample is selected on metallicity only, and
the RR Lyrae variables are selected via their variability, we have no
kinematic selection effects in the sample. RR Lyraes are generally
thought to be very old, while the giants and RHB stars could possibly
have a larger range of age, including both younger and older objects. The RR
Lyrae sample has a somewhat higher median metallicity than the giant
and RHB sample because the latter was identified with objective-prism
searches which tend to favor extremely weak-lined stars. 

\subsection{Angular Momentum plots}

\begin{figure}[h!]
\centering
\includegraphics[scale=0.3,angle=-90]{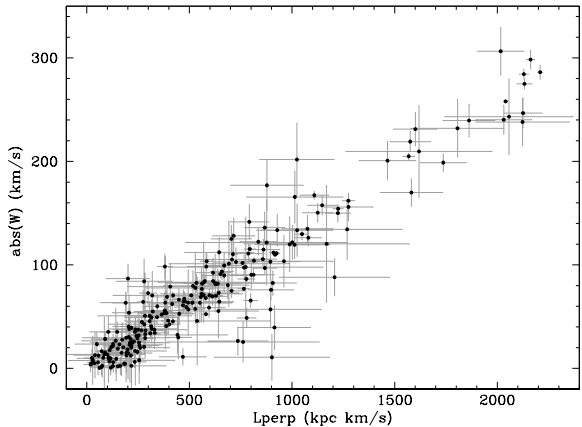}
\caption{Plot of angular momentum \lp\ vs the absolute value of the
  star's W velocity.}
\label{lperpvsw}
\end{figure}

\citet{amina_nature} used the angular momentum components \lz\ and \lp\ to
show that around 10\% of local halo stars were formed via destruction
of a satellite whose debris now occupies the inner halo. They compared
the amount of rotation of a given star's orbit (\lz) to \lp, which
combines the other two components of angular momentum (\lp =
$\sqrt(L_x^2 + L_y^2)$)\footnote{$L_x = yv_z - z v_y ; L_y =
zv_x - xv_z ; L_z = xv_y - y v_x$; see the Appendix of \citet{kepley07}
for a detailed discussion of our coordinate system.}. For a local
sample, \lp\ is dominated by $v_z$, the velocity perpendicular to the
plane, since $y$ and $z$ are close to zero, while $x$ is around 8
kpc. Figure \ref{lperpvsw} shows this clearly. Thus the \lz,\lp\ diagram is similar to the $v_\phi$ vs $v_z$
diagrams which contrast rotational angular momentum with the vertical extent of
an orbit. Stars with \lz\ less than zero are on retrograde orbits.

\begin{figure*}
\centering
\includegraphics[scale=0.5,angle=-90]{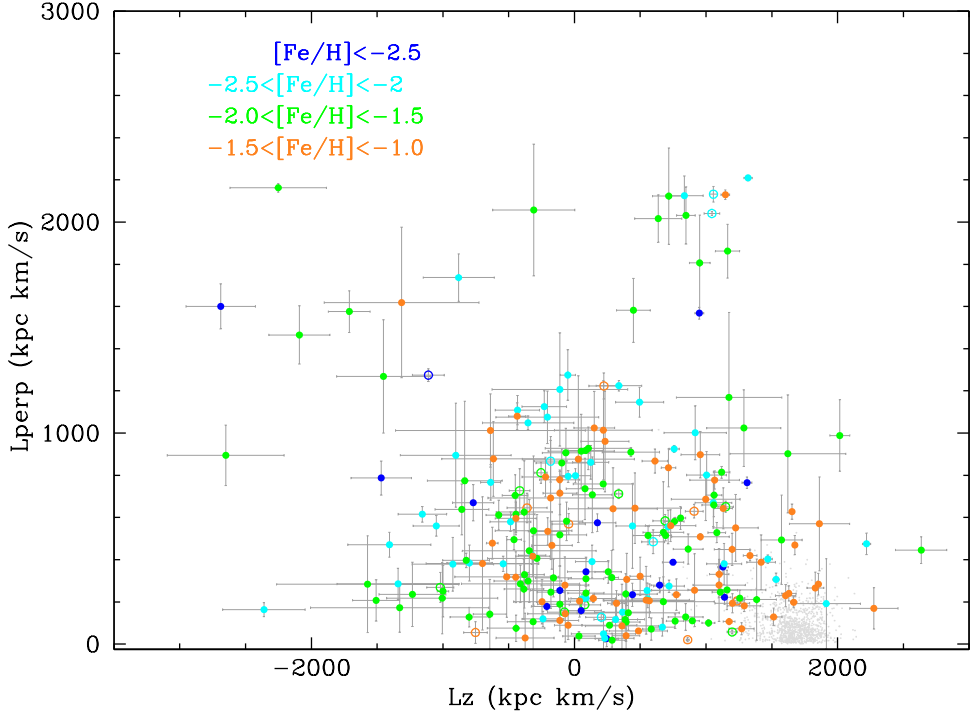}
\caption{Plot of angular momenta \lz\ (measuring rotation) vs \lp\ (in
  our sample, this correlates with the distance a star's orbit reaches
  from the disk plane) for stars with [Fe/H]$\leq -1$. It can be seen
  that the errors on these quantities are quite small for most of the
  sample. RHB stars are plotted with open circles. Stars from
  \citet{nordstrom} with [Fe/H]$>$--1.0 (which are predominantly from
  the disk and thick disk) are plotted in grey, and the globular
  cluster M4 is shown with an open 5-pointed star.
\label{lzlp}}

\end{figure*}

Figure \ref{lzlp} shows the distribution of stars in our sample in \lz\
and \lp.  To show the location of stars with near-circular orbits
which never reach far from the plane, the stars from the
\citet{nordstrom} sample, a magnitude-limited sample of stars with
distances up to $\sim$200 pc, are shown in grey on the same
diagram. The Nordstr{\"o}m et al.\ sample is almost completely
dominated by thin disk stars. As expected for a sample of stars in the
thin disk whose orbits are confined to the plane, few of these stars
have \lp\ greater than 200 kpc km s$^{-1}$, and their \lz\ values are
higher than almost all stars in our halo sample.

\begin{figure}[h]
\centering
\includegraphics[scale=0.45]{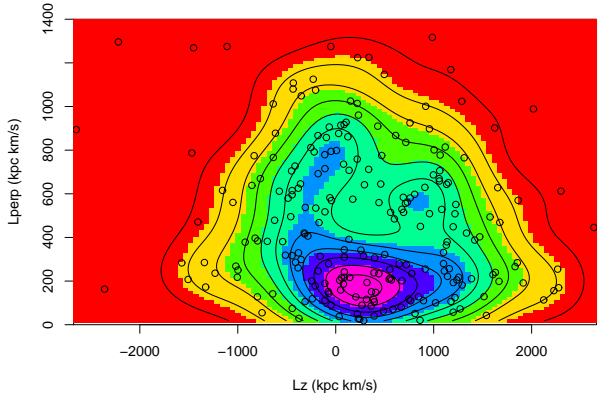}
\caption{Kernel Density Estimate plot of angular momenta \lz\ vs \lp\
for stars in our sample with [Fe/H] $\leq -1.0$. }
\label{lzlp_contour}
\end{figure}

\citet{kepley07} have discussed star streams in the sample: outliers
such as the \citet{amina_nature} group near (\lz,\lp)=(1000,2000) kpc
km s$^{-1}$ and the very retrograde stars with (\lz,\lp) near (--2000,1500)
kpc km s$^{-1}$.  In this paper we focus, not on these outliers, but on the
majority of the stars in the sample which occupy the region normally
described as the smooth, well-mixed halo. Thus, for the rest of the
paper, we consider stars with \lp $ < 1400$ kpc km s$^{-1}$. Figure 
\ref{lzlp_contour} shows a contour plot of the stellar density in this
diagram.

The classical description of halo kinematics uses a velocity ellipsoid
\citep{swar} with independent Gaussian distributions in $v_x,v_y$ and
$v_z$. In the small range of distance in our sample (recall that the
median distance is 1 kpc) these assumptions will lead to roughly
Gaussian distributions of \lz\ and \lp\ as well.  It is clear from
Figure \ref{lzlp} that such a velocity ellipsoid is not a good
description of our sample, even when outliers are removed by
restricting it to stars with \lp $<$ 1400 kpc km s$^{-1}$. The only part of
the sample that appears smooth and close to Gaussian is the region
occupied by stars whose orbits are quite confined to the plane: those
with low \lp\ ($\lesssim$ 400 kpc km s$^{-1}$). This is the region of \lp\
occupied by disk and thick disk stars in Figure \ref{lzlp}.  For
values of \lp\ above 400 kpc km s$^{-1}$, the distribution looks quite
clumpy. For example, there is a pronounced gap around (\lz,\lp) =
(500,400) kpc km s$^{-1}$ and clumps at (750,500) and (0,900) kpc km s$^{-1}$.

\subsection{Statistical Analysis}

We use two statistical approaches to quantify the structure in the
(\lz,\lp) diagram. In the first approach, which we present here in
detail, we fit mixtures of Gaussian distributions to
\lz\ distributions for different ranges of \lp, decide how many
components are needed in each \lp\ range (based on statistical model
selection procedures), and then estimate the value of \lp\ where the
distributions change character. Then we compare the results of the
first mixture fits with those based on the actual estimated ``change
points'' (the values of \lp\ where the distributions change
character).

We begin by dividing the sample by eye into three ranges of \lp:
\mbox{(0--350,350--700,700--1400)} and determining whether each
distribution of \lz\ is best described by a single Gaussian:

$$f\sim N(\mu,\sigma^2)\qquad$$

\noindent or a mixture of two:

$$ f\sim pN(\mu_1,\sigma^2_1)+(1-p)N(\mu_2,\sigma^2_2), 0<p<1$$

\begin{figure}[h]
\centering
\includegraphics[scale=0.35]{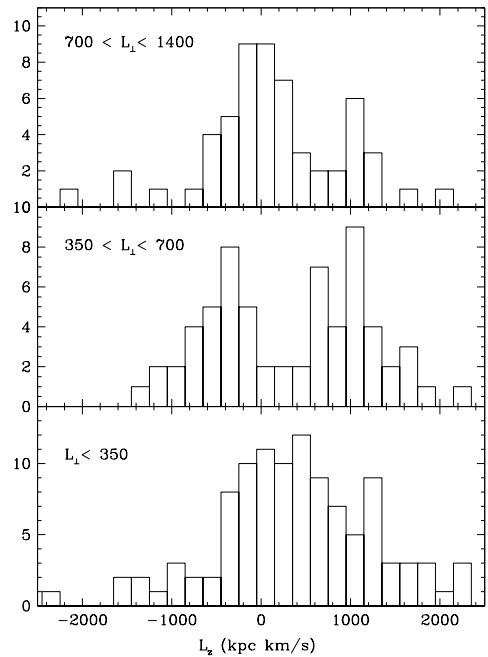}
\caption{Histograms of \lz\ for stars with [Fe/H] $\leq -1.0$, for three
  ranges of \lp\ as shown. }
\label{lzhists}
\end{figure}

Figure \ref{lzhists} shows the three histograms. It can be seen that
the histogram for the lowest \lp\ values (stars whose orbits are
confined to the disk plane) looks unimodal, while the other two (stars
with orbits which reach further from the plane) appear less so.  

We decided whether to use a unimodal or bimodal model using an
extension to the standard procedure of calculating the likelihood of
the sample under a given model, the Akaike
Information Criterion \citep[][AIC hereafter]{akaike}:

$$AIC=-2\log L(\hat{\theta})+2d$$

\noindent Here $L(\hat{\theta})$ is the likelihood function for the model with
parameters $\theta$ and $d$ is the number of parameters in a competing
model, unimodal or bimodal. A model with one extra parameter is thus
penalized by the factor of $d$ when comparing values of
log(likelihood).

Model comparisons using the AIC can be seen in Table \ref{llk_aic},
where the preferred models are shown in bold face.  A competitive
model selection criterion is Schwartz' Bayesian Information Criterion
\citep[BIC, ][]{schwarz} which changes $2d$ to $log(n).d$, where $n$
is the number of data points. If we had used the BIC for the \lp = 0
-- 350 range, it would have favored the one-component model even more
strongly, because this range of \lp\ contains the largest number of
stars. However, \citet{hemant} have demonstrated that AIC is a
better criterion for comparison of mixture models, so we prefer to use
this for our analysis. It should be noted that although the difference
between the AIC values for one and two components looks small compared
to its large value, a difference of 2 in its numerical value is
equivalent to the amount which justifies an additional parameter;
also, different analyses discussed below give similar results.

\begin{deluxetable}{rrll}
\tabletypesize{\scriptsize}
\tablecaption{Model Comparison for data grouped in \lp\ by
  eye \label{llk_aic}}
\tablewidth{0pt}
\tablehead{
\colhead{\lp\ range} & \colhead{n} & \colhead{AIC(1)}&\colhead{AIC(2)} 
}
\startdata
0--350    &103 & {\bf 1685.431}&1686.274 \\
350--700  &63 &1041.645&{\bf 1035.361} \\
700--1400 &61 &1018.468&{\bf 1015.161} \\
\enddata

\end{deluxetable}

We have shown that when we divide the sample into three \lp\ ranges by
eye, we find that the stars whose orbits are confined near the plane
have a different distribution than those whose orbits reach further
from the plane: the former have a simple Gaussian distribution of \lz,
while the latter have a bimodal distribution.  We can now remove the
effect of our subjective choice of \lp\ range using a change point
analysis \citep{cso97,chej00,ganocy}, which will solve for the best
values of \lp\ to separate the three groups. In particular we are
interested in the value of \lp\ where the distribution changes from
unimodal to bimodal, and in interpreting this number in terms of the
Galaxy's known stellar populations.

For the change point analysis we fit the following model to the data,
solving for the two dividing points $L_{\perp,1}$ and $L_{\perp,2}$:

$$f_1(L_z | L_\perp < L_{\perp,1}) \sim N(\mu_1,\sigma_1^2)$$
$$f_2(L_z | L_{\perp,1}\leq L_\perp < L_{\perp,2})\sim p N(\mu_2,\sigma^2_2)+(1-p)N(\mu_3,\sigma^2_3)$$
$$f_3(L_z | L_{\perp,2}\leq L_\perp <1400)\sim  N(\mu_4,\sigma_4^2)$$,
where $0<p<1$.

(We have chosen a single Gaussian for the highest \lp\ group to make
the change point analysis simpler; note that the middle group has the
strongest deviation from Gaussian shape.)

The dividing points which produce the highest likelihood are
$(L_{\perp,1},L_{\perp,2})$ = (349,709) kpc km s$^{-1}$; these are
quite close to the original values chosen by eye. Repeating the
analysis summarized in Table \ref{llk_aic} with these slightly
different values produces the same result: the sophisticated AIC
criterion prefers a Gaussian distribution for \lp\ less than 349 kpc
km s$^{-1}$, and mixtures of two Gaussians for the other two ranges of
\lp.

The lower change point, \lp = 349 kpc km s$^{-1}$, is particularly
interesting because it represents the value of \lp\ below which the
distribution of \lz\ is Gaussian. A rough calculation (setting $y=z=0$
and $x=8$) shows that this \lp\ ``change point'' corresponds to $|v_z|
\sim$ 40 km s$^{-1}$. The $z$ velocity dispersion of all stars in our
sample with \lp $<$ 350 kpc km s$^{-1}$ is 30 km
s$^{-1}$,\footnote{Fortuitously, these two quantities do not vary
  exactly in lockstep, which would mean that truncating \lp\ would
  produce a trucated distribution in $v_z$ and make estimation of the
  velocity dispersion difficult.} intermediate between the thin disk
($\sigma_z = 20$ km s$^{-1}$) and the thick disk ($\sigma_z = 40$ km
s$^{-1}$).

Table \ref{musig} shows the best fit mean and sigma for these
models. First, it should be noted that none of the groups have
rotational properties at all close to the thin and thick disks'
(\meanlz $\sim$ 1700 kpc km s$^{-1}$). The low \lp\ group (stars with orbits
closely confined to the plane) shows a mild prograde rotation
(\meanlz = 364 kpc km s$^{-1}$, corresponding roughly to $v_\phi$ = 45
km s$^{-1}$). The other two groups (containing stars with orbits reaching
further from the plane) have more complex distributions of Lz, as
can be seen in Figure \ref{lzhists}. Lz is roughly zero in the mean but quite clumpy, making the mean a less informative parameter. 

\begin{deluxetable*}{rrrrrr}
\tablewidth{0pt}
\tablecolumns{6}

\tablecaption{Parameters of best fit Gaussian mixtures for \lz\
  distributions for  different \lp\ ranges\label{musig}}
\tablehead{
\colhead{\lp\ range} & \colhead{$p$} & \colhead{$\mu_1$} &
\colhead{$\sigma_1$}  & \colhead{$\mu_2$} &\colhead{$\sigma_2$} }
\startdata
0--350   & 1.0 & 364 & 848 & \nodata & \nodata \\
350--700 & 0.45 & --460 & 348 & 1088 & 567 \\ 
700--1400 & 0.70 & 186 & 1035 & 54 & 194 \\
\enddata
\end{deluxetable*}

It is important to note here that the somewhat higher mean rotation of the
group with orbits very close to the plane is not due to thick disk
contamination: Figure \ref{lzlp} shows that there are almost no stars
in this lowest \lp\ bin in the region occupied by disk and thick disk
stars, which have near circular orbits. The low \lp\ stars in our
sample are on highly eccentric orbits close to the plane, as was first
noted by \citet{jesper}. The presence of this low scale height
component of the metal poor halo was not found by previous studies,
perhaps because of the significantly larger errors or the
high-latitude sampling of some surveys which led to a lack of data
close to the plane.

What is the spatial distribution of these stars with low \lp? We
cannot measure it directly from our sample, because there were
latitude-related selection effects in the initial surveys on which it
was based (see Section 3.2 below), and our sample contains only stars
from these surveys which have been followed up extremely thoroughly by
a number of authors, thus giving very accurate kinematical
measurements, rather than all stars identified down to a certain
magnitude by the surveys.  It is also not possible to simply compare
the $z$ velocity dispersion of these stars with that of the thick
disk, because the thick disk's large degree of rotational support adds
extra flattening. However, we can make an estimate of the axial ratio
of this component using its velocity ellipsoid, and employing the tensor
virial theorem.  \citet{bnt} discuss the case of a self-gravitating
population in the context of the flattening and rotation of elliptical
galaxies, and \citet{amina_review} discusses the case of a tracer
population such as the stellar halo. We find that the velocity ellipsoid for
stars with \lp\ less than 350 kpc km s$^{-1}$, ($\sigma_U,\sigma_V,\sigma_W$)
= (144,111,30), corresponds to a flattening $c/a$ =
0.2.  This contrasts markedly with the flattening for the entire halo
at the solar radius ($c/a$=0.6), which can either be derived directly
from the models of, for example, \citet{preston91} or from the local
velocity ellipsoid \citep[see][]{amina_review}.

\begin{figure}[h!]
\centering
\includegraphics[scale=0.33]{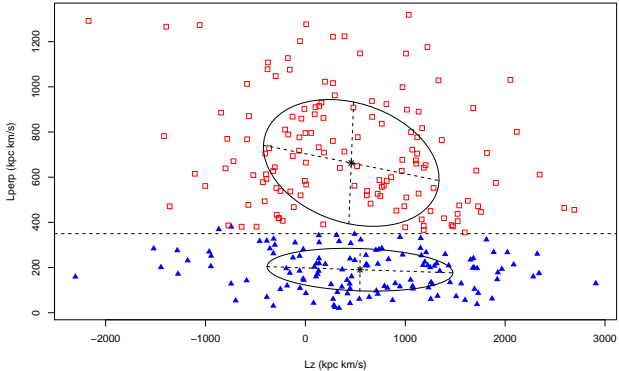}
\caption{MCLUST fit of elliptical Gaussians to our data. It can be
  seen that this alternative analysis produces very similar answers:
  a highly flattened elliptical Gaussian for the low values of \lp,
  separated from the other data near \lp = 350 kpc km s$^{-1}$.}
\label{rongfang}
\end{figure}

In our second statistical analysis, we used the model-based clustering
method in the R project for statistical
computing\footnote{http://www.r-project.org} package
MCLUST\footnote{http://cran.r-project.org/doc/packages/mclust.pdf}. In
the MCLUST analysis, the data are fit by a mixture of a finite number
of two-dimensional Gaussian distributions, and the best fit is
computed using a modification of the AIC with a second order
correction for small sample sizes
(AICc\footnote{http://en.wikipedia.org/wiki/Akaike\_information\_criterion}).
We considered both 2-D Gaussians with major and minor axes parallel to
the \lz\ and \lp\ axes, and more complex models where the Gaussians
could be oriented in any direction.  A good fit was obtained with two
elliptical Gaussians. The one at low \lp\ is quite flattened, with
major axis parallel to the \lz\ axis; the other, for the higher
\lp\ values, is a rounder elliptical Gaussian. These Gaussians can be
seen in Figure \ref{rongfang}. The separation point between the two
elliptical Gaussians is close to \lz = 350, confirming, with a
different analysis technique, the results of the change point analysis
reported above. Another analysis, using the AIC only, divided the data
with \lp\ greater than 350 into five different clusters, so we should
not take the two-Gaussian result shown in Figure \ref{rongfang} to
indicate a lack of substructure in the data with high \lp.

\subsection{Binding Energy}

\begin{figure}[h!]
\centering
\includegraphics[scale=0.4]{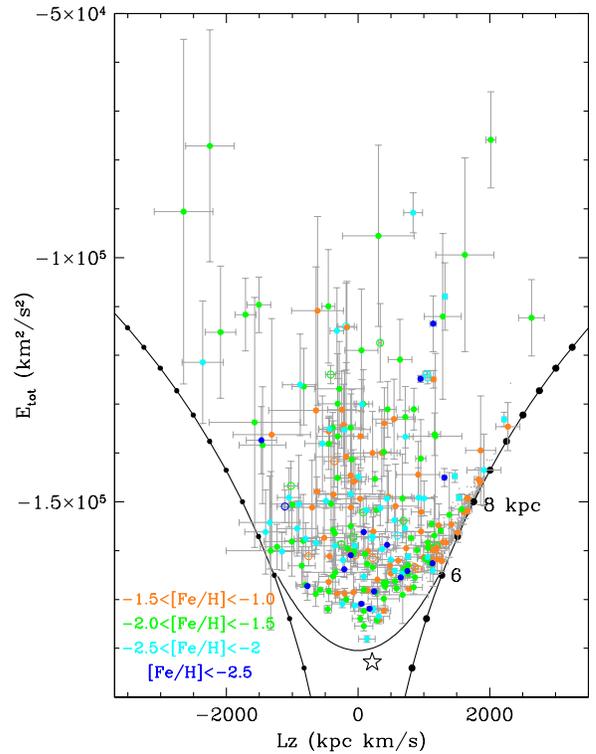}
\caption{The energy-angular momentum diagram for our sample. Points
  are coded by metallicity as in Figure \ref{lzlp}, and the
  predominantly disk stars from \citet{nordstrom} are shown in
  grey. RHB stars are again plotted with open circles.
  Black lines with closed circles show the circular orbits at a given
  position, with orbits at 6 and 8 kpc marked, while the parabola at
  the bottom shows the position of orbits in the plane with
  apogalacticon 7 kpc. The star is the globular cluster M4, which has
  a near-radial orbit and is currently 6.3 kpc from the galactic
  center.}
\label{lzet}
\end{figure}

It will be useful in our discussion of the different processes which
deposit stars in the inner halo to consider the total energy of each
star, as well as its angular momentum.  Figure \ref{lzet} shows the
traditional energy-angular momentum plot. (Note that we have computed
these values for the entire sample, not just those stars with
\lp\ less than 1400 kpc km s$^{-1}$.) To calculate total energy we
have used the Galaxy potential of \citet{kathryn_potential} with a
somewhat smaller disk mass of $7.5 \times 10^{10}$ \msun and a dark
halo circular velocity of 155 \kms. We set the potential energy equal
to zero at R=200 kpc.

It can be seen that the majority of stars in our sample have higher
binding energy (lower $E_{tot}$) than the energy corresponding to the
LSR velocity. This is expected in a sample of halo stars because of
the halo's steep density distribution \citep[$R^{-3.5}$, measured for
  RR Lyraes by][]{vivas06}.  The clumpiness that we see in the
\lz,\lp\ diagram is also visible in this diagram, particularly along
the lines 
\lz=1200 and 0 kpc km s$^{-1}$. The clump near \lz=1100 kpc km
s$^{-1}$,\etot=--160,000 km$^2$/s$^2$ has energy and angular momentum
similar to the Arcturus group \citep{eggen98,julio_arcturus}. We also
see the ``plume'' at \lz$\sim-300$ kpc \kms and a large range of
$E_{tot}$ which has been remarked upon by \citet{dana02} and
\citet{brook03}. They suggested that this ``plume'' is debris from the
progenitor of $\omega$ Centauri.

\begin{figure*}
\centering
\includegraphics[scale=0.5,angle=270]{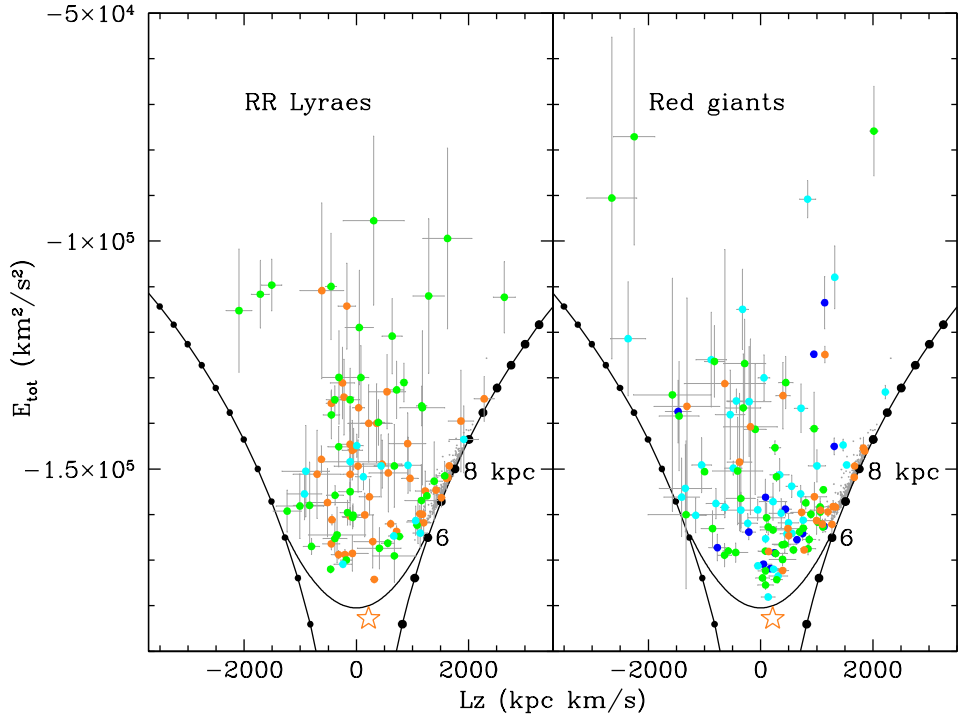}
\caption{The energy-angular momentum diagram for our sample, showing
  RR Lyrae variables on the left and (first ascent) red giants on the
  right. Points are coded by metallicity as in Figure
  \ref{lzlp}. Lines mark loci of circular orbits and apogalacticon
  7 kpc, as in Figure \ref{lzet}.  The orange star is the globular
  cluster M4, which has [Fe/H]=--1.22 \citep{kraftivans}.\label{lzetbystartype} }

\end{figure*}

\begin{figure}[h!]
\centering
\includegraphics[scale=0.3,angle=270]{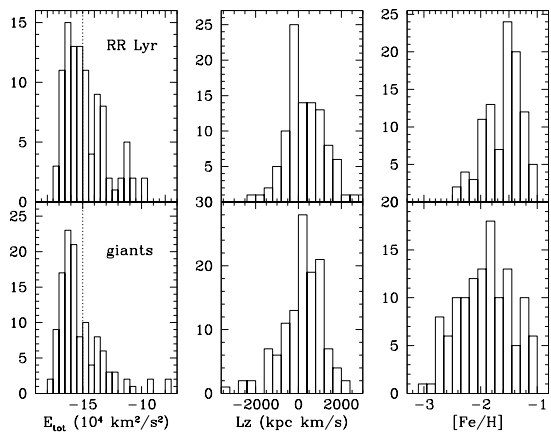}
\caption{Histograms of \etot\, \lz\ and [Fe/H] for the RR Lyrae and
  red giant subsamples. RHB stars have been omitted. The energy of the
  LSR orbit is shown on the left panels with a dotted line. It can be
  seen that the giants are more tightly bound on average than the RR
  Lyraes.  }
\label{rrg_ethists}
\end{figure}

In Figures \ref{lzetbystartype} and \ref{rrg_ethists} we see the
intriguing result that the RR Lyraes and red giants in our sample show
differences in binding energy\footnote{The differences between the two
  samples in the \lz--\lp\ diagram are much more subtle.}.  (Note that
we have excluded the RHB stars from this comparison for clarity.) The
red giants (which have significantly more metal-poor stars) are on
average more tightly bound to the Galaxy than the RR Lyraes. We find
that 55\% of the RR Lyrae sample have binding energy less than the
LSR's, compared with only one third of the red giants. We tested the
statistical significance of this result by sampling the (122 star) red
giant sample with replacement 1000 times in order to produce samples
of the same size as the RR Lyrae sample (99 stars). None of the
thousand samples had more than 55\% with binding energy less than the
LSR's, and only one had half the sample with less binding energy than
the LSR. It is clear that the giants and the RR Lyrae variables have
different distributions in binding energy. We note that since less
than 10\% of our sample are RHB stars, it is not possible to make a
similar check for these stars.

This result is a kinematic counterpart of the \citet{preston91} result that
RR Lyraes and BHB stars have different spatial distributions, with the
RR Lyraes less centrally concentrated than the BHB stars.  It is also
related to the different population properties of both field RR Lyraes
and globular clusters when they are divided according to the
Oosterhoff dichotomy \citep[eg][]{leecarney,miceli} which is likely
related to age \citep{leecarney,jurcsik}.

The giants in our sample will evolve onto the horizontal branch as
they age, and since the horizontal branch morphology in the solar
neighborhood is very blue \citep{kinsuntkraft}, most of them will
become BHB stars.  Thus we see that the lower binding energy of our RR
Lyraes is consistent with the result of \citet{preston91} that RR
Lyraes have a shallower spatial distribution than BHB stars, and so
are more likely to be found in the outer halo. Our RR Lyrae sample
contains more objects from the outer halo than the red giant sample,
although not all RR Lyraes in our sample are outer halo objects, as
can be seen in Figure \ref{rrg_ethists}.

Connecting horizontal-branch morphology to age would suggest that
these outer halo objects were also younger, as originally suggested by
\citet{sz} and then by \citet{zinn93}. The younger age of an accreted
outer halo compared with the inner halo formed {\it in situ} makes
sense in the light of the theoretical expectation that the first stars
to form in the Galaxy would be in the rare high-density peaks, and
would be centrally concentrated \citep{diemand05}; outer halo objects
would form somewhat later. The dominance of retrograde orbits in the
low binding-energy stars of both types is also interesting in the
light of the discovery of \citet{daniela} that the outer halo has
retrograde rotation. It will be very interesting to see whether the
nearby BHB stars (which may be older than the RR Lyraes) share the
properties of the red giants in our sample in this diagram.

There are other interesting differences between the RR Lyrae and red
giant kinematics in our sample: many of the more bound red giants are
have prograde orbits, including a number close to the circular orbit
line, while the RR Lyraes have a smoother distribution of orbital
angular momenta.

\subsection{Trends with Metallicity}

Trends with metallicity are interesting because they give a handle on
two processes: dissipational collapse with star formation subsequently
enriching later generations, and dynamical friction, which is more
effective for massive objects which produce stars with higher mean
metallicity because of the mass-metallicity relation
\citep[eg][]{henrylee}. We see an interesting difference in kinematics
between the very metal-poor stars (with $\fe < -1.5$) and the
metal-richer halo stars (with $-1.5 < \fe < -1.0$), in Figure
\ref{lzlp}, where stars with different metallicity are shown in
different colors.  While the entire sample has values of \lz\ ranging
from --3000 to 2700 kpc km s$^{-1}$, there are no stars with intermediate
metallicity ($-1.0> \fe > -1.5$) which have \lz\ below --800 kpc
km s$^{-1}$. 
\citep[Note that with a median halo metallicity of --1.6, eg][, these intermediate-metallicity stars account
for almost half of the halo, although they are under-represented in
our sample because of the relative ease of identifying very metal-poor
giant stars in surveys.]{ryannorris}

\begin{figure}[h!]
\centering
\includegraphics[scale=0.4]{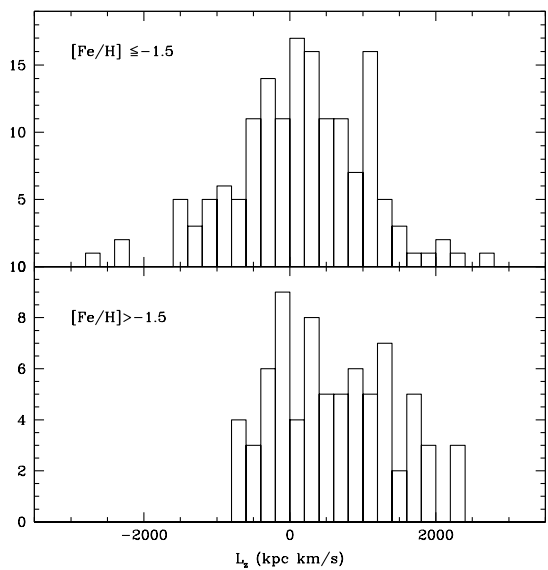}
\caption{Histograms of \lz\ values for stars with \lp\ less than 1400
  kpc km s$^{-1}$. Upper panel shows stars with [Fe/H] $<-1.5$, lower panel
  those with [Fe/H] between --1.0 and --1.5. }
\label{lzbyfe}
\end{figure}

We see the trend with metallicity more easily in the histograms of Figure
\ref{lzbyfe}. The median \lz\ value for the metal-richer stars is 545
kpc km s$^{-1}$, significantly higher than the median for the metal-poor
stars (135 kpc km s$^{-1}$).  The sample of \citet{jesper} also shows this
difference in \lz\ for more metal-rich halo stars.

\section{Discussion}

\subsection{Halo flattening: another component}

The first suggestions that the halo's flattening might be variable
with radius were in the 1950s and 60s by \citet{schmidt} and
\citet{kinman65}, studying globular clusters and RR Lyraes
respectively. While both studies were contaminated by some
metal-richer objects, it is unlikely that this was the only reason for
their findings.  \citet{hartwick} suggested a two-component halo based
on RR Lyrae samples.  \citet{zinn93} also found a two-component halo
by studying the kinematics, space distribution, metallicity and ages
of globular clusters in the Galaxy.  The Old Halo clusters of the
inner halo show a flattened distribution and have some rotation
(although much less than the disk's). The Younger Halo is spherical
and has no net rotation.  More recently, \citet{miceli} discussed a
two-component halo with distinct radial distributions for different
Oosterhoof types in their large sample of RR Lyraes, and
\citet{daniela} used a very large sample of halo stars from the SDSS
survey to extend the description of the two-component halo. Of
particular importance to an understanding of the population
characteristics of the two components is Carollo et al's demonstration
that the near-spherical outer halo has a net retrograde rotation and
that it is more metal-poor than the flattened inner halo.

All of the field star results are for relatively distant samples.
What of samples like our own, with many stars less than 1 kpc from the
plane?  \citet{jesper} used an early local sample of 151 metal-poor,
predominantly giant and RR Lyrae stars with less accurate data than
ours.  Using only stars with $\fe <$ --1.5, they reconstructed the
density distribution of the halo, found that these metal-poor stars
had a two-component structure, and noted that their flattened shape
was due to velocity anisotropy, not rotation.  While they do not
explicitly quote the axial ratio of the flattened component, it can be
seen in their Figure 6 that it is quite flat: $c/a$ around 0.25. In
general, our analysis agrees well with these early
results. \citet{jesper} suggested that either the metal-poor stars in
the flattened component formed in an anisotropic, dissipative collapse
(with the anisotropic infall caused by a flattened dark halo) or by
late infall of metal-poor gas clouds which ``plunged into the disk''
and formed stars there.

\citet{preston91} used BHB stars and RR Lyraes (which were dominated
by stars more than 2 kpc from the plane except for some very close to
the galactic center) and derived a density law for the halo which
became more spherical with increasing radius, with axial ratio $c/a$ =
0.5 at the galactic center. However, when they used this model to
predict the RR Lyrae density near the Sun, they found that it
under-predicted the known value by a factor of two, concluding that
``there exists a metal-poor population, confined to a volume near the
galactic plane, which has remained undetected in intermediate and
high-latitude surveys.'' It seems very likely that Preston et al were
describing what we call the low \lp\ component of the halo.
\citet{greenmorrison} also suggested the existence of this highly
flattened halo component using a sample of BHB stars within 500 pc of
the Sun.

Thus there are {\it two} flattened components in the inner
halo: the moderately flattened, $c/a \sim 0.6$ component identified by
\citet{kinman65,hartwick,preston91,zinn93} and \citet{kinsuntkraft} from more
distant samples of halo stars, and the highly flattened, $c/a \sim
0.2$ component foreshadowed by \citet{jesper}, \citet{preston91} and
\citet{greenmorrison} and revealed in more detail in this work. We
will discuss possible origins for these two components below.

\citet{chibayoshii} and \citet{chibabeers} also studied local samples,
although the one used by the latter (the sample of Beers et al. 2000)
suffered from large distance errors for a number of stars \citep[see
  Section 2.1 and][]{kepley07}. \citet{chibayoshii} focused their
attention on quantifying the contribution of the metal weak thick disk
to their sample. Curiously, although \citet{chibabeers} use a local
sample, they conclude that their results are similar to those of the
more distant samples such as \citet{preston91}: a halo with an axial
ratio decreasing as R decreases and only moderate flattening in the
center. We suggest that the large distance errors of some of the stars
in their sample might have made it difficult to distinguish between a
moderately flattened and a highly flattened inner halo. We explored
the effects of a larger measurement error using our data, by adding a
random number picked from a Gaussian with twice the error in \lz\ and
\lp\ to each datapoint. Much of the structure that we see is no longer
visible, showing that our careful sample selection was needed.

Recently, \citet{warrenbrown} have presented a preliminary kinematic
analysis of a large sample of 2414 BHB stars with distances larger
than 2 kpc from the galactic plane. They find a flattened component,
but identify it with the metal-weak thick disk because of its
significant rotation. However, in their paper they simply solve for mean
rotation of the entire sample as a function of distance from the
plane, and find both a trend with $z$ and a surprisingly large random
variation in mean rotation (from +150 to --60 km s$^{-1}$) over the range of
$z$ heights covered by their sample. \citet{kinman08} have shown,
using proper motion information for the most nearby stars in this
sample, plus another sample of nearby BHB stars, that the mean
rotation of BHB stars close to the plane is close to zero. Thus the
flattened component in the \citet{warrenbrown} sample is likely to be
associated with our low \lp\ group.

It is also interesting to note that the nearest globular cluster, M4,
has properties consistent with membership in the highly flattened
halo. Using the data given in \citet{dinescu99} we find that it has
$z$=--0.5 kpc, \lz = 215 $\pm$ 141 kpc km s$^{-1}$ and \lp = 32 $\pm$
22 kpc km s$^{-1}$. We show its position in the energy and angular
momentum diagrams of Figures \ref{lzlp} and \ref{lzet} with an open
5-pointed star.  Using the data from \citet{dinescu99} and
\citet{dinescu07}, we find that there are another four globular
clusters with $R_{gc}$ between 6 and 10 kpc which have similar orbital
properties to our low \lp\ group: NGC 4372, 4833, 5139 ($\omega$
Centauri) and 6779, and another four with $R_{gc}$ less than 6 kpc:
NGC 5986, 6093, 6656 and 6712. 

\subsection{The thick disk}

It is remarkable how little contribution the thick disk makes to our
sample; the region of the \lz--\lp\ diagram occupied by the
\citet{nordstrom} local dwarfs, a sample almost totally dominated by
the disk, contains less than 5\% of our sample. Given the fact that
thick disk stars outnumber halo stars by 1--2 orders of magnitude in
our region, this could put strong constraints on the size of the
low-metallicity tail of the thick disk metallicity
distribution. Earlier claims of a significant low-metallicity
tail \citep{mff} were over-estimates because of problems with the DDO
metallicity calibration \citep{twarogs_ddo,sean,chibayoshii}.

\begin{figure}[h!]
\centering
\includegraphics[scale=0.25]{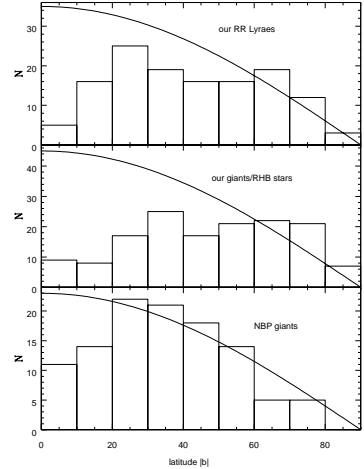}
\caption{Histograms of latitude $|b|$ for (top) the RR Lyraes in our
  sample (middle) the red giants and RHB stars in our sample, and
  (bottom) the giants of \citet{nbp}. The curve on each plot is
  cos(b). It can be seen by comparison with the curve on each plot
 that our local sample has serious incompleteness
  at low latitude, and the NBP sample is much more complete. }
\label{completeness}
\end{figure}

However, both the RR Lyrae and the red giant/RHB samples are seriously
incomplete at low latitudes, as can be seen in Figure
\ref{completeness}. This incompleteness would reduce the number of
thick disk stars found in our sample. While this is unlikely to be a
strong effect for our very local sample (recall that the mean distance
of our stars is 1 kpc), we cannot use this sample to obtain an
unbiased estimate of the density of metal-weak thick disk stars unless
we engage in complex corrections for our selection effects. By
contrast, the sample of \citet{nbp} which was used for first studies
of the metal-weak thick disk \citep{nbp,mff} shows much less
incompleteness with latitude, as can be seen in the bottom panel of
Figure 6. An upcoming analysis of the metallicity of the stars in this
sample \citep[Beers, private communication,][]{krugler04} would be
the best way to obtain an unbiased estimate of the local density of
the metal-weak thick disk.

\subsection{Summary of Results}

Pulling together all of the above analysis, we have the following results.

 (1) We find evidence for an additional component to the inner halo,
which is highly flattened ($c/a \sim 0.2$), pressure-supported and has
a small prograde rotation. This component is not associated with the
metal-weak thick disk, which is predominantly rotationally
supported. The new component shows a smooth distribution of \lz.

This component is separate from the moderately flattened inner halo
which was discovered using samples of halo stars located further than
1 kpc from the Sun, for example by
\citet{kinman65,hartwick} and \citet{preston91}.  This moderately flattened
component has axial ratio ($c/a$) of $\sim$0.6. If the globular
clusters are representative of the field stars in this moderately
flattened component, it is predominantly old and also has a small
prograde rotation. \citet{daniela} also ascribe a ``modest prograde
rotation'' to their moderately flattened inner halo.

The highly flattened component forms about 40\% of our
sample. However, this could be an under-estimate of its true density
as the samples on which we base our analysis avoid low latitudes (see
Figure \ref{completeness}).

 (2) The remaining 60\% of stars in our sample, whose orbits
reach further from the plane, have a mean rotation close to zero and a
clumpy distribution in angular momentum and energy.

 (3) We see trends with metallicity. The low metallicity stars in our
sample ($\fe <-1.5$) show a mean rotation near zero and a clumpy
distribution in energy and angular momentum.

The intermediate-metallicity stars ($-1.5 < \fe < -1.0$) cover the
whole range of \lp\ (in other words, they belong both to the highly
flattened component and to the other(s)), but there is a higher
proportion of these intermediate-metallicity stars in the low \lp\
region. The intermediate-metallicity stars have a small prograde
rotation and a clumpy distribution in angular momentum and energy.

(4) The RR Lyrae variables in our sample have a lower binding energy,
on average, than the red giants, with half of the RR Lyraes being less
tightly bound than the LSR orbit, in contrast to the red giants where
only one third of the stars are less tightly bound. Because the local
horizontal branch morphology is very blue, this suggests that the
younger RR Lyrae stars are less strongly bound to the Galaxy. The RR
Lyrae sample also appears to contain more stars on retrograde orbits.

\section{Formation and evolution of the inner halo}

We will now discuss the different stages of halo formation and the
constraints placed on various theories by our data. In particular we
will examine possible formation paths for the highly flattened halo
component which is supported by velocity anisotropy.

\subsection{The First Halo Stars}

Theory predicts that the first stars to form in the Galaxy's halo
occupy dark halos which formed from very rare high-density peaks in
the primordial density field.  At this early stage (perhaps beginning
earlier than z=10, less than 0.5 Gyr after the Big Bang) the Galaxy's
disk had not formed.  \citet{diemand05} use simulations to predict
that the density distribution of such stars is more centrally
concentrated than the rest of the halo.  Many of these first stars will
now be found in the inner few kpc of the Galaxy; however there will
still be a few with orbits that reach the solar neighborhood.

How might we identify these first stars?  We would expect the very
first stars to have extremely low metallicity. However, the amount of
chemical enrichment that occurs in the earliest star formation in the
halo is not well constrained, so we do not know whether only extremely
metal-poor stars were formed at this stage, or whether some stars
closer to typical halo metallicities were also formed.  We see no
unique pattern in the angular momentum or energy distribution of the
most metal-poor stars in our sample. We plan to extend our sample to
stars of extremely low metal abundance and check whether their
kinematics are different.

Where would the early halo stars, formed long before the disk, appear
in the \lz--\lp\ diagram?  We would expect them to have low total
energy (be tightly bound to the Galaxy) and have both low $L$ and low
\lz. Interestingly, the very metal-poor star CD--38 245 ([Fe/H]=--4.0)
is in the original sample of \citet{att94} but was excluded from our
sample because of its distance (3.3 kpc). However, its angular
momentum and energy are well-determined, and fit the expectation for a
star formed very early: \lz = --110 kpc km s$^{-1}$, \lp = 254 kpc km
s$^{-1}$ and \etot = --1.6 $\times 10^5$ km$^2$s$^{-2}$: it is on a
near-radial orbit, confined to the inner halo.  It is quite  possible that
CD--38$^\circ$ 245 was formed early, and is a genuine ``original
inhabitant'' of the inner halo. 

Because the Galaxy itself is quite small at such early times, the
first halo stars to form would have low total angular momentum, and
thus low \lp. However, they cannot form our entire low \lp\ group,
because of its large range in \lz. Also, it is unlikely that the
chaotic conditions under which the first stars formed would allow a
very flattened configuration, such as this group, to survive. This is
also the reason why any disk which formed during the early period
where the Galaxy's halo was growing via violent merging would not
survive.

We should note, however, that there is another evolutionary path which
will also produce stars with low angular momentum and energy:
dynamical friction dragging satellites in from the outer halo.  So the
possession of low angular momentum and low energy are not sufficient
to identify early halo stars.

\subsection{Accretion and halo growth before disk formation}

\subsubsection{Substructure in E and L} 

Before the disk forms, mergers will cause the Galaxy's gravitational
potential to vary strongly. This violent relaxation might smooth out
any dynamical record of the origins of stars formed then, and lead to
a kinematically smooth distribution. This has been the source of a
general expectation that the inner halo be smoother than the outer
halo. However, in our inner halo sample, the only stars which display
a smooth \lz\ distribution are ones which occupy the highly flattened
component. We have argued above that we would not expect this
component to be formed before the disk because such a flattened
spatial distribution would be destroyed by mergers. Thus its
smoothness cannot be ascribed to violent relaxation and is not
relevant here.

However, we would expect many of the stars in the remainder of the
sample to be formed at this stage of the halo's growth, and we see no
evidence for a complete smoothing of kinematical quantities by violent
relaxation there. We are not aware of any other observational study
which finds such a clumpy distribution in energy and angular momentum.
We suggest that the larger distance errors of previous samples (which
become magnified when quantities such as energy and angular momentum are
measured) have obscured the intrinsic clumpiness of the data there.
We note that our clumpy distribution in angular momentum and energy
does not contradict the result of \citet{gould} that no more than 5\%
of the local halo comes from a given stream because multiple streams
can come from a single progenitor. Results from simulations of the
dark halo \citep{amina02,amina03,diemand05,diemand07} have a similar
result: the velocity distribution of the dark matter is quite smooth
there, despite a small number of progenitors supplying the field stars
of the inner halo.

The clumpy distribution in energy and angular momentum implies that
either the violent relaxation was not violent enough to destroy
structure in angular momentum and energy, or this process was limited
to the inner few kpc of the Galaxy, and did not reach the solar
neighborhood. This is good news for galactic archaeology, as it will
allow us to trace various halo substructures to very early times.  If,
in addition to the preservation of some initial conditions through the
violent relaxation stage, a small number of accreted objects formed
most of the stars in the inner halo, as suggested by the simulations
of \citet{amina02} and \citet{amina03}, this would give a natural
explanation for the clumpiness we see. It is unlikely that such a
clumpy distribution in angular momentum would be produced by a large
number of progenitors with distinct orbital properties.

\subsection{Pure accretion}

The corollary of the expectation that the earliest stars will form in
the high-density peaks near the center of the Galaxy is that star
formation in the lower-density regions that join the Galaxy as its
halo grows should start somewhat later. This leads to the expectation
that RR Lyrae samples, which should be younger on average than BHB
samples, should be less strongly bound to the Milky Way. We have
discussed the fact that we see this in our sample in Section 2.4.

Now let us focus on ``pure'' accretion, unassisted by dynamical
friction, which builds up the outer halo, and ask whether any of these
outer halo objects would be found in our local sample. As the Galaxy
grows, and its turnaround radius (where objects separate from the
Hubble flow and become bound to the Galaxy) increases, it becomes
harder for accreted objects to reach the solar neighborhood unless
they are on near-radial orbits.  However, radial orbits are quite
common for infalling satellites: \citet{benson} shows that more than
half of the orbits of satellites in the simulations he studies have
eccentricity between 0.9 and 1.0 when they enter their host halo,
showing that this is not an impossible
requirement\footnote{\citet{benson} cautions that larger studies are
  needed to disentangle the effects of redshift and mass on this
  quantity.}.

Such objects, with near-radial orbits, have all components of angular
momentum near zero. If dynamical friction has not occurred, then we
have the additional diagnostic that the orbits will have low binding
energy (high total energy) as well, since they fell into the Galaxy
from a large distance.  Stars from these late-accreted small objects
will be found in the region near \lz\ $\sim$ 0 and \etot\ greater
than, say, --1.4$\times 10^5$ km$^2$ sec$^{-2}$.  Intriguingly, there
is a preference for slightly retrograde orbits in this region of the
\lz--\etot\ diagram, and we see more of this component in the
(possibly younger) RR Lyrae sample than in the red giants.  This may
be related to the result of \citet{daniela} that the outer halo has a
small net retrograde motion; if the material which was added to the
Milky Way at later times had a different mean angular momentum from
the material which formed the disk, then the only part of this outer
halo that could reach the solar neighborhood without help from
dynamical friction would be the stars on near-radial orbits. The
remainder of these stars would then populate the outer halo. However,
it must be remembered that dynamical friction can also populate this
area, and in fact many of these stars have also been claimed to be the
debris of $\omega$ Cen's progenitor \citep{dana02,brook03,meza}.

We have noted that it is not likely that our highly flattened halo
component formed before the disk because the ongoing mergers which
only stopped when the present-day disk formed would tend to heat this
very flattened component. However, the {\it moderately} flattened
inner halo could have been formed during mergers of gas-rich infalling
objects at early times, in a number of ways. For example, when
gas-rich infalling objects merge, the gas is likely to form stars with
low net angular momentum. This process is referred to as ``dissipative
merging'' by \citet{bekkichiba} and seen (to excess) in early
cosmologically-based galaxy formation simulations such as that of
\citet{nfw95}. Stars formed in this way would likely have low
metallicity and make up part of the inner halo, although the gas might
have experienced some pre-enrichment. \citet{bekkichiba} use this
process to explain the moderately-flattened, somewhat metal-enriched
inner halo.

\subsection{Dynamical Friction}

There is another way for an accreted satellite to sink to the inner
halo, even without a radial orbit.  Dynamical friction is a purely
gravitational process, caused by the formation of a gravitational wake
as a massive object travels through a density field. It allows
satellites, given sufficient mass and time before disruption, to
transfer energy and angular momentum to the dark halo and sink toward
the galactic center and the disk plane.  Dynamical friction from the
halo will permit massive satellites which are accreted at a later time
to end up in the inner halo, even if their initial perigalactica are
large.  The efficiency of dynamical friction is a strong function of
the mass ratio between object and host. In the case of a single host,
it varies strongly with the satellite mass \citep[$M^2$;][]{bnt}; see
also Figure 13 of \citet{walkermihos}, so we
would expect the more massive objects \citep[which are also likely to
form more metal-rich stars:][]{marioaraa,tremonti,henrylee,erb} to be
more successful in reaching the solar neighborhood via this process.

The strong dependence of dynamical friction on the mass of the
accreted object is important for the inner halo: it means that it will
be dominated by debris from only a few relatively massive satellites
\citep{amina02,amina03}, and may well be related to the clumpiness we see
in E and L. \citet{meza} present simulations to show that the
disruption of a single satellite via dynamical friction can in fact
lead to a number of long-lived groupings in energy and angular
momentum space. 

Dynamical friction can also affect the spatial distribution of the
inner halo, as follows.  Theory predicts that only a few progenitors
contributed most of the mass of the inner halo, making it less likely
that the result of these few accretions will be exactly
spherical. Accretion of hundreds of smaller sub-halos will likely
produce a spherical outer halo, while it would not be surprising if
the stars stripped from the few massive objects which formed the inner
halo produced a somewhat flattened inner stellar halo. So the
cosmological predictions for the growth of the dark halo will quite
naturally produce a moderately flattened inner halo and a
near-spherical outer halo.

\subsection{Disk Formation}

As we have noted above, the Galaxy's potential needs to settle down
(the epoch of major mergers needs to end) before we can form either
the dynamically cold thin disk or our highly flattened halo component,
since they would be destroyed by these violent variations in the
gravitational potential. The need for angular momentum conservation in
producing the large disks that we see today also requires this. 
While the classical picture of smooth gas dissipation and infall
conserves angular momentum, mergers between gas-rich objects at early
times can transfer both energy and angular momentum to the dark halo
\citep[eg][]{nfw95}. Angular momentum will not be conserved during this
process, which dominates at early times, unless feedback processes
keep gas out of the cores of the merging objects. This was first
suggested by \citet{nfw95}; we will see below that recent simulations
have confirmed the result.

Galaxy formation simulations cannot trace star formation in detail
because the spatial resolution needed far exceeds what is
computationally possible when studying an entire galaxy.  Thus average
prescriptions are used to model the complex, multi-phase gas physics.
In many of the early simulations of galaxy formation, the angular
momentum of the gas was not conserved: gas cooled, collapsed and
formed stars early, and then mergers transferred its angular momentum
to the dark halo \citep[eg][]{steinmetzmuller,abadi03} and only small
disks were formed (the ``angular momentum problem''). Recent
simulations which include a more complex treatment of feedback
\citep[eg][]{springelhernquist} have formed larger, more realistic
late-type disks \citep{robertson,zavala,governato07}.  Because the
accretion of a large satellite will destroy a dynamically cold stellar
disk, the present-day disk starts to form only after major mergers
have ceased.  However, the star formation rate in the disks formed in
even the most recent studies tends to peak at early times, soon after
the last merger \citep[see, for example, ][]{abadi03,governato07}, in
contrast to the star formation rate in the Milky Way disk, which is
thought to have been roughly constant for most of its life
\citep{twarog80,rocha2000}. More work on modelling feedback and gas flows is
clearly needed before the major properties of late-type disks are
reproduced.

In the simulations of \citet{abadi03} and \citet{springelhernquist05},
the last major mergers before disk formation were between gas-rich
galaxies. It is possible that one or both of the halo's flattened
components were created in a way that was intimately related to the
formation of the disk.  In these simulations, the stars which were
already part of the galaxies in the last major mergers formed a
flattened component, and gas from these merging objects then settled
into a thin disk and formed stars. (It is not clear at this stage
whether the flattened component formed from accreted stars is caused
by the specific initial conditions of these simulations or whether it
is a general outcome.) \citet{abadi03b} identify this flattened
component (despite its rather large scale height of 2.7 kpc) with the
thick disk, which has a scale height of $\sim$1 kpc in the Milky Way
\citep{reylerobin,ng} and note that it has kinematics ``intermediate
between halo and disk'': the stars have a mean rotation of around 150
km s$^{-1}$ but a large velocity dispersion.  This moderately
flattened component might also be related to the flattened inner halo
first suggested by \citet{kinman65}.  As more simulations are
published, it will be interesting to see the variation of scale height
and rotation of the components formed from the stars in the satellites
which define the orientation of the disk.

It is also possible that the low \lp\ group was formed at this stage.
The gas that came in with them could form stars in a flattened
configuration that would form our low \lp\ group, instead of starting
to form the disk. Which outcome occurs (rotationally supported disk or
highly flattened halo) will depend on the angular momentum content of
this gas. Since, by construction, the merger of these satellites was
the last major one experienced by the Galaxy, any flattened stellar
configuration formed after this time will not be subjected to huge
variations in the Galaxy's potential, and so it will be able to
survive. This would explain the fact that the highly flattened
component has stars which are more metal-rich than the overall halo,
because stars from the galaxies which merged would have already
polluted their ISM with some metals. While gas on predominantly radial
orbits in an extremely flattened configuration (like the present-day
young thin disk) will be likely to ``self-cross'' and shock, forming a
more rotationally supported component, the low \lp\ group does not
have such extreme flattening, so the gas carried in with the
satellites which experience the last major mergers could still form
the low \lp\ component.

The accretion and settling to the disk plane of gas and subsequent
star formation will continue to build the thin disk in a process that
continues to this day. Orbits of inner halo stars will be affected by
the potential of this highly flattened disk, and could increase the
flattening of the inner halo via adiabatic contraction, as originally
pointed out by \citet{binneymay}. \citet{chibabeers01} have shown that
this process is not sufficient to form an inner halo with axial ratio
as high as 0.7; we have seen that the actual flattening of the inner
halo is closer to 0.6. So adiabatic contraction alone is unlikely to
form the moderately flattened inner halo.

\subsection{Fashionably late: after disk formation}

After the disk forms, we have another source of dynamical friction
which can affect the orbits of objects which reach the inner
Galaxy. Unlike dynamical friction with the halo, this process will be
more efficient for objects on prograde orbits because the lower
relative velocity of satellite and disk stars enhances the formation
of the gravitational wake and so increases dynamical friction
\citep{walkermihos,abadi03,meza}.  We would thus expect to see debris
from objects which experience dynamical friction with the disk to tend
to have a net prograde rotation. It is still possible for dynamical
friction to bring objects on retrograde orbits into the inner Galaxy,
but the process will be significantly slower. For example,
\citet{bekkifreeman} show how the angular momenta of stars in a
possible progenitor of the retrograde, tightly bound globular cluster
$\omega$ Centauri change as it is dragged down into the plane.

After the merger rate of the early Galaxy settles down and the disk
forms at around $z$=2, the physical conditions are appropriate for
the survival of our flattened (low \lp) group. 
As time goes on, the turnaround radius continues to grow, and, as
noted above, it becomes steadily more difficult for objects to reach
the inner halo without the help of dynamical friction. Satellites on
extremely radial orbits can still do so, but will not form our low
\lp\ component because it has quite a range in \lz. So, dynamical
friction is an important part of the origin of this component.
This fits in well with the result shown in Figure \ref{lzbyfe}: more
massive objects will also form more metal-rich stars on average, so
this gives a natural explanation of the higher metallicity of this
component.

Can we constrain the number of progenitors whose accretion formed this
highly flattened component? The stars in this component have quite a
large range of energy and \lz, although their \lp\ range is small.
The energy and angular momenta of stars lost from a satellite will be
determined by both the orbital energy and angular momentum of the
satellite itself, and the position of the stars in the Galaxy when
they are stripped. If the disruption happens within a few kpc of the Galaxy's
center, the satellite's finite physical size will ensure that stars will
gain different amounts of potential energy and angular momentum
depending on their distance to the center when they are stripped from
the satellite. Thus it is possible that even a single progenitor
produced this component, if it was disrupted close to the Galaxy's
center. It is also possible that more than one satellite was disrupted
to form it.

The formation of the disk will flatten the inner Galaxy's potential,
making it easier for dynamical friction with the halo to drag objects
down into the plane. There are now two options: first that the
satellite(s) were disrupted before they had a chance to experience
dynamical friction with the disk or thick disk; and second that they
were disrupted later, in which case dynamical friction with both halo
and disk would have occurred. If only dynamical friction with the halo
occurred, then the slightly prograde rotation of the highly flattened
component is simply a coincidence. This would be easier to explain
with one or a few progenitors.

If dynamical friction with the disk itself occurred, we have a
``built-in'' explanation for the small prograde rotation of this
component. Dynamical friction with the disk is more efficient for
objects on prograde orbits, since they will spend more time near the
gravitational influence of the disk stars. In this case we would
expect there to be a difference in \meanlz\ between objects with low
\lp\ which experienced dynamical friction with the disk, and those
whose orbits do not spend enough time close to the plane to experience
disk dynamical friction, and so would be found with higher \lp.
 
At what distance above the plane does dynamical friction from the disk
become important? We can produce a rough estimate for the current
epoch by calculating where the density of the present-day thin disk
and dark halo are roughly equal in the solar neighborhood. Using a
simple isothermal halo with $V_c$ = 220 km s$^{-1}$ and $r_c$ = 2 kpc and
assuming a thin disk stellar density of 0.1 \msun\ per pc$^3$ we find
that the densities are roughly equal at 2.3 thin disk scale heights
($z$ = 800 pc) So we would expect stars to need to spend significant
time within 1 kpc of the plane in order to feel dynamical friction
from the disk. This is quite interesting in the light of the \lp\ =
350 result, which is equivalent to a population scale height
similar to the thick disk's. 

Why should dynamical friction produce a {\bf smoother}
distribution for stars close to the plane? The more massive satellites
would have a larger stellar velocity dispersion, but they will also
produce more streams, which would need velocity accuracy of order 1 km
s$^{-1}$ to separate. If there was a single progenitor, the range of
potential energy and angular momentum imparted to stars as they were
stripped at different galactocentric radii would likely produce such a
smooth distribution. It is more difficult to explain this smoothness
in the case of a larger number of progenitors.  

Another possibility is that the halo star orbits with low \lp\ are
close enough to the plane to feel the effect of inhomogeneities in the
disk potential due to spiral arms, giant molecular clouds and the
bar. Scattering and heating caused by interactions with these massive
objects would smooth the velocity distribution of the stars. However,
the non-circular orbits of these halo stars ensure that the stars
would not linger long within the gravitational reach of these large
density inhomogeneities.

The origin of this low \lp\ group from processes involving dynamical
friction makes the nine globular clusters with extended horizontal
branches and orbits similar to the low \lp\ stars particularly
interesting. \citet{youngwook07} suggest that the clusters with
extended horizontal branches also have helium-enhanced
second-generation subpopulations (which are directly observed in
several cases, notably in $\omega$ Centauri). If the clusters were
originally cores of ancient dwarf galaxies which have since disrupted
\citep{bekkifreeman,bekkinorris}, and the galaxies were sufficiently
massive to be affected by dynamical friction, then it would not be
surprising to find them well represented in our low \lp\ group. In
fact, five out of the six most tightly bound clusters with these
orbital properties have extended horizontal branches, so their orbital
properties are consistent with their origins as cores of ancient dwarfs.

In summary, we feel that late accretion, and dynamical friction with
the dark matter halo (and perhaps also existing disk and thick disk
stars) is the best way to explain the origin of the highly flattened
inner halo.

\section{Summary}

Using our sample of metal-poor giants, RHB stars and RR Lyraes, we
find that the metal-poor halo has a highly flattened component, with
axial ratio $c/a \sim 0.2$, a similar scale height to the thick disk,
in the neighborhood of the Sun.  This component was suggested by
\citet{jesper} and predicted by \citet{preston91} and
\citet{greenmorrison}. The highly flattened component has a small
prograde rotation, but most of its stars are on fairly eccentric
orbits; this small rotation does not affect its flattening
significantly. Stars of only moderate metal deficiency ($-1.5 < \fe <
-1.0$) are more likely to be found in this component. This component
is distinct from both the metal weak thick disk, which is primarily
rotationally supported, and from the moderately flattened inner halo
detected by many previous workers, which has a much less flattened
density distribution.

We suggest that the most likely formation paths for this component
would be ``fashionably late'' ones: either the stars formed from the
gas accreted in satellites whose merger preceded the formation
of the disk, or the stars were carried into the inner Galaxy by
dynamical friction acting on their massive progenitor. Other formation paths
such as dissipative mergers at early times could have formed the
moderately-flattened halo, but are unlikely to have formed our highly
flattened component because the violent merging activity at these
early epochs would have destroyed such a coherent component.

We find that the other stars in our sample, whose orbits reach further
from the plane, exhibit quite a clumpy distribution in energy and
angular momentum, suggesting that the chaotic conditions under which
the early halo formed were not violent enough to erase the record of
the conditions under which they were formed. Any truly smooth inner
halo is either a very small component or only exists inside the solar
neighborhood. This clumpy structure also suggests that only a small
number of objects contributed the majority of stars to the inner halo,
in line with theoretical expectations. It is possible that the
progenitor galaxy of $\omega$ Centauri was one of these objects.

In addition we find that the RR Lyrae variables in our sample are
somewhat less tightly bound to the Galaxy, on average, than the red
giants, and show a more retrograde rotation. This is consistent with
earlier claims of a younger outer halo such as that of \citet{sz} and
\citet{zinn93}, with the retrograde outer halo of \citet{daniela} and
with the difference in kinematics between RR Lyrae and BHB stars at
the NGP detected by \citet{kinman07}.

\section{Acknowledgements}

AAK was supported by a NSF Graduate Research Fellowship during
portions of this work; HLM by NSF grants AST-0098435 and AST-0607518;
JS was supported by a grant from the NSF. AH acknowledges financial
support from NOVA and NWO. HLM would like to thank Chris Mihos,
Kathryn Johnston and James Bullock for useful discussions, Joanne
Tidwell for suggesting the paper title, and Monty Python's Flying
Circus for support.

\end{document}